\documentclass{article}

\usepackage[latin1]{inputenc}
\usepackage[english]{babel}

\usepackage{amsmath, amsthm, amsfonts, amssymb}
\usepackage{graphicx}
\usepackage{color}
\usepackage[affil-it]{authblk} 
\usepackage{epstopdf}
\usepackage[lofdepth,lotdepth]{subfig}

\newcommand{\norm}[1]{\left\Vert#1\right\Vert}

\newcommand{\mf}{\mathbf}

\newcommand{\mfs}{\mathbf \Sigma}
\newcommand{\mfr}{\boldsymbol \rho}

\DeclareMathOperator{\tr}{tr}

\title{Maximum Likelihood Estimation of the correlation parameters for elliptical copulas}
\author{Lorenzo Hern\'andez \thanks{lorenzo.hernandez@qrr.es}}
\author{Jorge Tejero \thanks{jorge.tejero@qrr.es}}
\author{Jaime Vinuesa \thanks{jaime.vinuesa@qrr.es}}
\affil{Quantitative Risk Research S.L. Madrid, Spain}

\begin{document}
\maketitle

\begin{abstract}
\noindent We present an algorithm to obtain the maximum likelihood estimates of the correlation parameters of elliptical copulas. 
Previously existing methods for this task were either fast but only approximate or exact but very time-consuming, especially for high-dimensional problems. 
Our proposal combines the advantages of both, since it obtains the exact estimates and its performance makes it suitable for most practical applications. 
The algorithm is given with explicit expressions for the Gaussian and Student's t copulas. \\


\end{abstract}

\section{Introduction}
Copulas are a popular statistical tool to describe the dependence 
between two or more random variables. They allow to model a multivariate 
distribution in a flexible way, by describing the marginals and the 
dependence structure separately. There are many available families
of parametric copulas that represent different types of dependence 
structures and are described by parameters that control their strength and form.
Their use has grown extensively in the past two decades, especially in 
the field of financial mathematics. The reader may refer to 
\cite{AB11}, \cite{BK08}, \cite{EHJ03}, \cite{HMM13}, \cite{JRR04} or \cite{W10} 
for some references in the area.\\

\noindent Specifically, if $F_{\mathbf{X}}(\mathbf{x})$ is the multivariate distribution function of a $d$-dimensional random vector $\mathbf{X}$ with continuous marginal distributions $F_{X_i}(x)$, then the (unique) copula defined by this distribution is given by Sklar's theorem
\begin{equation}
    C(\mathbf{u}) = F_{\mathbf{X}}\big(F_{X_1}^{-1}(u_1),...,F_{X_d}^{-1}(u_d)\big)\, , \, \, \mathbf{u} \in [0,1]^d. 
\end{equation} 
where $F_{X_i}^{-1}$ is the inverse distribution of the $i$-th marginal.  $C(\mathbf{u})$ is the distribution function of the copula, that encodes the dependence structure of the random vector $\mathbf{X}$. It has uniform margins and does not depend on the particular form of the marginal distributions of $\mathbf{X}$. When it exists, the corresponding copula density is defined by $c(\mathbf{u}) = \frac{\partial}{\partial u_1} ... \frac{\partial}{\partial u_d} C(\mathbf{u})$. For more details on copulas the reader can refer, for example, to \cite{N99}.\\

\noindent Elliptical copulas are the underlying copulas of multivariate elliptical distributions. The probability density function of these distributions, when it is defined, can be expressed as 
\begin{equation}
f_{\mathbf{X}}(\mathbf{x};\boldsymbol{\mu},\boldsymbol{\Sigma},\psi) = K \psi\Big( (\mathbf{x}-\boldsymbol{\mu})^{\top} \boldsymbol{\Sigma}^{-1}(\mathbf{x}-\boldsymbol{\mu})\Big)\, , \, \,  \mathbf{x}\in\mathbb{R}^d ,
\end{equation}
where $\psi$ is a non-negative function on $\mathbb{R}^+$ with an appropriate integrability condition\footnote{In order to define a valid density, the function $\psi$ must satisfy the condition $\int_0^{\infty} \psi(r^2)r^{d-1}dr < \infty$.}, $K$ is the normalization constant, $\boldsymbol{\mu}$ is a location vector and $\boldsymbol{\Sigma}$ a dispersion matrix with the properties of a covariance matrix (i.e. symmetric and positive-definite). The reader may refer to \cite{CHS81} for a discussion on elliptical distributions.\\

\noindent The densities of elliptical copulas, that arise from these distributions, do not depend on the location parameters and depend on the dispersion parameters only through the scaled matrix, $\boldsymbol{\rho}$, given by
\begin{equation}
    \mfr_{ij} = \frac{\mfs_{ij}}{\sqrt{\mfs_{ii}\Sigma_{jj}}},
\end{equation}
that is, they just depend on the correlation matrix associated to $\boldsymbol{\Sigma}$. Note, in particular, that the copula density, $c(\mathbf{u};\boldsymbol{\rho},\psi)$, does not depend on the location and dispersion parameters that describe the marginals.\\

\noindent The copulas of the elliptical family are tractable and have straightforward simulation procedures. Due to this, they are extensively used, especially the Gaussian and the Student's t copulas, corresponding to the multivariate normal and Student's t distributions respectively. The density of the Gaussian copula is
\begin{equation}
c_{\text{Gaussian}}(\mathbf{u};\mathbf{\rho}) = \frac{1}{\sqrt{\vert \boldsymbol{\rho}\vert }} \frac{e^{-\frac{1}{2}\mathbf{g}^{\top}\boldsymbol{\rho}^{-1}\mathbf{g}}}{\prod_{i=1}^d e^{-\frac{1}{2} g_i^2} },
\end{equation}
where $\mathbf{g} = \{g_i=\Phi^{-1}(u_i) \}_{i=1}^d $, $\Phi$ being the standard univariate normal distribution function. For the Student's t copula, the density is
\begin{equation}
c_{\text{t}}(\mathbf{u};\mathbf{\rho},\nu) = \frac{\Gamma(\frac{\nu+d}{2})\Gamma(\frac{\nu}{2})^{d-1}}{ \sqrt{\vert \boldsymbol{\rho}\vert}  \Gamma(\frac{\nu+1}{2})^d}\frac{\left(1+\frac{\mathbf{s}^{\top}\boldsymbol{\rho}^{-1}\mathbf{s}}{\nu}\right)^{-\frac{\nu+d}{2}}}{\prod_{i=1}^d\left( 1+\frac{s_i^2}{\nu}\right)^{-\frac{\nu+1}{2}}} ,
\end{equation}
where $\nu$ is the degrees-of-freedom parameter, $\mathbf{s} = \{s_i=t_{\nu}^{-1}(u_i) \}_{i=1}^d$ and $t_{\nu}$ is the univariate Student's t distribution function. Note that the Gaussian copula is the limiting case of the Student's t copula as $\nu \rightarrow\infty$. We stress again the fact that the previous expressions define valid copulas only if $\boldsymbol{\rho}$ is a correlation matrix; as using a more general covariance matrix would result in non-uniform margins.\\

\noindent There are two widely used approaches to estimate the correlation parameters 
of elliptical copulas from data: the \textit{method of moments}, 
which is based on matching empirical and theoretical rank correlation measures, and the \textit{maximum likelihood method} (MLM), with which this work is concerned, based on the maximization of the joint probability density 
of the sample. For a detailed description of the method of moments the reader may refer to \cite{MFE05}.\\

\noindent The implementation of the MLM requires, in general, numerical optimization techniques, since closed-form 
expressions for the estimates do not always exist. This is especially relevant when 
dealing with high-dimensional elliptical copulas, since the number of parameters in the correlation matrix grows as the square of the number of dimensions and standard optimization methods cannot be practically applied for such problems.\\

\noindent In the present article we introduce an efficient procedure to obtain the maximum likelihood estimates of the correlation parameters $\mfr$ of elliptical copulas. Other parameters, for instance the degrees-of-freedom parameter in the case of the Student's t copula, will be assumed given. Note that, in that particular case, using a one-dimensional optimization routine in conjunction with the presented algorithm would allow the efficient estimation of all the parameters of the Student's t copula.\\

\noindent When focusing on elliptical copulas with density, given a sample $U=\{ \mathbf{u}_t\}_{t=1}^n$, with $\mathbf{u}_t = \{ u_{t,i} \in [0,1] \}_{i=1}^d$, the MLM method amounts to solving the following constrained maximization problem
\begin{equation}
    \widehat{\mfr} = \underset{\mfr}{\operatorname{argmax}} \Big\lbrace L(\mfr),\, \, \, \big \vert \, \, \, \mfr \in \mathcal{P} \Big\rbrace  
\end{equation}
for the log-likelihood
\begin{equation}
 L(\mfr) = \sum_{t=1}^n \log c(\mathbf{u}_t;\mfr)
\end{equation}
where $\mathcal{P}$ is the space of correlation matrices, that is, the space of all symmetric, positive-definite matrices with diagonal elements equal to one.\\

\noindent In section 2 we describe a new algorithm to obtain the MLM estimator of the correlation parameters of elliptical copulas. In section 3 we provide test results for the Student's t copula, comparing our algorithm with other existing estimation methods. The conclusions will be presented in section 4 and, finally, the explicit algorithms for the Student's t and the Gaussian copulas are detailed in the appendix.\\

\section{Derivation of the algorithm}

\noindent To our best knowledge, there are no previous methods to obtain the exact maximum likelihood estimates of the correlation parameters of elliptical copulas efficiently. There are, however, widely used approximate methods for both the Gaussian and the Student's t copulas. In essence, these methods are based on finding a solution $\widehat{\mfs}$ to the problem in a less constrained space, $\mathcal{C}$, the 
space of symmetric positive-definite matrices, and then projecting this solution 
to $\mathcal{P}$ using the projector
\begin{eqnarray}
\Pi\, : \,  \mathcal{C}    & \longrightarrow & \mathcal{P}\nonumber\\
     \Pi(\widehat{\mfs}) & \longrightarrow & \mf{A} \widehat{\mfs} \mf{A} ,
\end{eqnarray}
where $\mf{A}_{ij} = \frac{\delta_{ij}}{\sqrt{\mfs_{ii}}}$ and $\delta_{ij}$ is the Kronecker delta. In particular, for the Gaussian copula the maximization problem in $\mathcal{C}$ has an exact solution (see, for example \cite{MFE05}, section 5.5.3)
\begin{equation}
\label{GAUSS_APPROX}
\hat{\mfs} = \frac{1}{n} \sum_{t=1}^n \textbf{g}_t  \textbf{g}_t^{\top}\, , \, \, g_{t,i} = \Phi^{-1} (u_{t,i}).
\end{equation}
For the Student's t copula, from the critical point condition $\frac{\partial L}{\partial \mfr^{-1}}=0$, the following fixed-point iteration is proposed in \cite{BDNRR00}
\begin{equation}
\label{BOUYE_FP}
\widehat{\mfs}_{\left[ m + 1 \right]} = \left(1 + \frac{d}{\nu} \right) \frac{1}{T} \sum_{t=1}^T \frac{\mf{s}_t \mf{s}_t^\top}{\left( 1 + \frac{1}{\nu} \mf{s}_t^\top \widehat{\mfr}_{\left[ m \right]}^{-1} \mf{s}_t \right)}\, , \, \, s_{t,i} = t_{\nu}^{-1} (u_{t,i}),
\end{equation}
where the projection is performed at each iteration
\begin{equation}
\widehat{\mfr}_{\left[ m + 1 \right]} = \Pi( \widehat{\mfs}_{\left[ m + 1 \right]} ).
\end{equation}
Note, in particular, that the previous method is a particular case of this one when $\nu \rightarrow \infty$.\\

\noindent Although these methods are computationally efficient, the solutions obtained from them are not true maximizers of the likelihood function because, in general, the application of the projector $\Pi$ does not map a maximizer in $\mathcal{C}$ to a maximizer in $\mathcal{P}$. In fact, the error in the solutions can be significant, both in the likelihood and in the values of the parameters.\\

\noindent In order to address the constrained maximization, the basic idea of the algorithm presented in this work is to define a projected version of the log-likelihood function 
\begin{equation}
L^* = L \circ \Pi,
\end{equation}
and solve the maximization problem
\begin{equation}
\label{OPTIMIZATION_STAR}
    \widehat{\mfs} = \underset{\mfs}{\operatorname{argmax}} \Big\lbrace L^* (\mfs),\, \, \, \big \vert \, \, \, \mfs \in \mathcal{C} \Big\rbrace  ,
\end{equation}
so that the likelihood function is evaluated in a valid (correlation) parameter matrix. The copula correlation parameter estimate is obtained by the projection 
\begin{equation}
    \widehat{\mfr} = \Pi(\widehat{\mfs}).
\end{equation} 

\noindent Then, the necessary critical point condition for the projected log-likelihood that has to be satisfied by the solution of the maximization problem is\footnote{We write the critical point condition as the derivative with respect to $\mfs^{-1}$ instead of the derivative with respect to $\mfs$ because the log-likelihood depends on a quadratic form whose matrix is the inverse of the correlation matrix, and therefore the expression of the former derivative is simpler.}
\begin{equation}
\frac{\partial L^*(\mfs)}{\partial \mfs^{-1}} = \frac{\partial L(\Pi(\mfs))}{\partial \mfs^{-1}} =0.
\end{equation}
Using the chain rule and defining 
\begin{equation}
\mathcal{D}_{ij}(\mfr) = \frac{\partial L(\mfr)}{\partial \mfr^{-1}_{ij}},
\end{equation}
the condition can be written as
\begin{eqnarray}
    0 &=& \frac{\partial L^*(\mfs)}{\partial \mfs^{-1}_{ij}} = \sum_{kl}\frac{\partial \mfr^{-1}_{kl}}{\partial \mfs^{-1}_{ij}} \mathcal{D}_{kl}(\mfr) \nonumber\\
    &=&\sum\limits_{kl}\left(\delta_{ik}\delta_{jl}\sqrt{\mfs_{kk}\mfs_{ll}} -  \mfs^{-1}_{ij}\mfs_{ki}\mfs_{kj}\sqrt{\frac{\mfs_{ll}}{\mfs_{kk}}} \right) \mathcal{D}_{kl}(\mfr)
\end{eqnarray}
or, in matrix notation,
\begin{equation}
\label{IGDIRECTION}
0=\frac{\partial L^*(\mfs)}{\partial \mfs^{-1}} = \mf{A}^{-1} \Big( \mathcal{D}(\mfr) -  \mfr \ \text{diag} \left( \mathcal{D}(\mfr) \mfr^{-1} \right) \mfr \Big) \mf{A}^{-1} ,
\end{equation}
where $\text{diag}(\mf{X})_{ij} = \mf{X}_{ij} \delta_{ij}$. In order to solve the this equation, we use the fact that the critical point also satisfies
\begin{equation}
\mfs = \mfs - \lambda \frac{\partial L^*(\mfs)}{\partial \mfs^{-1}},
\end{equation}
which suggests the following fixed-point iteration scheme 
\begin{equation}
\label{INVERSE_GRADIENT}
\mfs_{[m+1]} = \mfs_{[m]} - \lambda \frac{\partial L^*(\mfs)}{\partial \mfs^{-1}} \Big{\vert}_{\mfs = \mfs_{[m]}}
\end{equation}
where $\lambda$ is a step size parameter small enough to ensure that $\mfs_{[m+1]}$ is positive-definite after the iteration.\\

\noindent The algorithm is implemented by using the corresponding log-likelihood derivative $\mathcal{D}(\mfr)$ for the particular copula to be estimated. The expressions for the Gaussian and Student's t copulas are given in the appendix, but in principle the algorithm is applicable to any elliptical copula for which $\mathcal{D}(\mfr)$ can be computed in closed form.\\

\noindent The iterative scheme $\eqref{INVERSE_GRADIENT}$ has some resemblance to a gradient ascent method but, instead of moving along the gradient direction (that in principle would seem optimal), it moves along the direction $\mathbf{V} = -\frac{\partial L^*(\mfs)}{\partial \mfs^{-1}}$. For that reason, we will denote it \textit{inverse gradient} algorithm\footnote{Note that this inverse algorithm is not a standard gradient ascent on the coordinates of the inverse matrix, as this would correspond to the iterative scheme $$\mfs_{[m+1]}^{-1} = \mfs_{[m]}^{-1} + \lambda \frac{\partial L^*(\mfs)}{\partial \mfs^{-1}} \Big{\vert}_{\mfs = \mfs_{[m]}}$$} hereafter. In fact, the directional derivative of $L^*$ in the direction $\mathbf{V}$ is positive. To prove this, note that, since   
\begin{equation}
    \frac{\partial L^*(\mfs)}{\partial \mfs^{-1}} = -\mfs \frac{\partial L^*(\mfs)}    
    {\partial \mfs} \mfs,
\end{equation}
the directional derivative of $L^*$ along this direction can be expressed as 
\begin{equation} 
\Delta_{\mathbf{V}} L^* \equiv \sum_{ij} \frac{\partial L^*(\mfs)}{\partial \mfs_{ij}} \frac{\mathbf{V}_{ij}}{\norm{\mathbf{V}}} = \frac{1}{\norm{\mathbf{V}}}\tr\left(\frac{\partial L^*(\mfs)}{\partial \mfs} \mfs  \frac{\partial L^*(\mfs)}{\partial \mfs} \mfs \right),
\end{equation}
\noindent where $\tr$ is the trace operator and $||\mathbf{V}||$ is the norm of $\mathbf{V}$. Since $\mfs$ is a real, symmetric, positive-definite matrix, it admits the decomposition $\mfs = \mathbf{O} \mathbf{Q} \mathbf{O^\top}$, where $\mathbf{O}$ is an orthogonal matrix and $\mathbf{Q}$ is a diagonal matrix with positive diagonal entries $\mathbf{Q}_{ii}>0$. Therefore, by defining $\mathbf{M} = \mathbf{O}^{\top}\frac{\partial L^*(\mfs)}{\partial \mfs}\mathbf{O}$ and using the cyclical property of the trace and the fact that $\mathbf{M}$ is symmetric, we have 
\begin{equation}
\Delta_{\mathbf{V}} L^* =\frac{1}{\norm{\mathbf{V}}} \tr \left( \mathbf{M} \mathbf{Q}\mathbf{M}\mathbf{Q}  \right) = \frac{1}{\norm{\mathbf{V}}}\sum_{ij} \mathbf{Q}_{ii}\mathbf{Q}_{jj} (\mathbf{M}_{ij})^2 \geq 0.
\end{equation}
\noindent In particular, this guarantees that, in each iteration of \eqref{INVERSE_GRADIENT}, $L^*$ increases for a sufficiently small value of $\lambda>0$.\\

\noindent Note that, if in the iterative scheme \eqref{INVERSE_GRADIENT} we replace the projected log-likelihood $L^*$ by $L$ and set $\lambda = \frac{2}{T}$, we recover the approximate method given in reference \cite{BDNRR00}, and therefore this value of $\lambda$ would seem a natural choice. However, for particular data samples this choice can produce matrices that are not positive-definite, and hence a smaller $\lambda$ is required in those cases.\\

\noindent Although we have not been able to provide a formal proof of the convergence of the algorithm, in all the performed numerical experiments, with a choice of $\lambda \lesssim \frac{1}{T}$, the algorithm not only converged but also outperformed the standard gradient method. For practical applications, we have found that using a simple adaptive scheme for the step size provides the best results in terms of performance. This adaptive scheme is described in the appendix.\\

\section{Numerical experiments}

\noindent In this section, we present the results obtained when comparing the proposed inverse gradient method with other existing estimation algorithms. On the one hand, we have compared the performance of the inverse gradient and other exact estimation algorithms that are included in widely used statistical software packages. On the other hand, we have also calculated the differences in likelihood between the solutions obtained by the inverse gradient and the method proposed in \cite{BDNRR00}, which we will denote \emph{approximate method} hereafter.\\

\noindent For the performance tests, we have used the exact maximum likelihood estimations available in R and Matlab\footnote{In R we have used the function \texttt{fitCopula} of package \texttt{copula}, with \texttt{method = "ml"} and, in Matlab, only for the Student's t copula, the method \texttt{copulafit} slightly modified to print the execution times of the estimations of the correlation matrices for fixed $\nu$ values.} and an implementation of the inverse gradient method written in Matlab. Though all the methods obtain practically the same solutions, the differences in execution times are very significant. Specifically, in several test cases with 100 observations of dimension 25, the inverse gradient converged always in less than 1 second while, for the Gaussian copula, R took always more than 1000 seconds and, for the Student's t copula, both R and Matlab took more than an hour in every test case.\\

\noindent For the Student's t copula, we have also compared the solutions given by the inverse gradient method with the ones obtained using approximate method. For this purpose, we have generated test cases with dimensions 2, 10 and 25, and $\nu \in \left\{1/2, 1, 2, 5, 10, 20, 50\right\}$. For each combination of these two parameters, $50000$ test cases have been generated. For each test case:

\begin{enumerate}
\item A random correlation matrix with the required dimension is generated. The eigenvalues of this matrix are independently sampled from a uniform distribution (and then renormalized so that their sum coincides with the dimension). A description of the algorithm used to generate these matrices can be found in \cite{DH00}, Algorithm 3.1.
\item 100 random vectors are sampled from a Student's t copula with $\nu$ degrees-of-freedom and correlation given by the generated matrix.
\item The correlation parameters of the copula are estimated from the obtained sample both with the inverse gradient and with the approximate method. In both cases, the parameter $\nu$ is fixed to the original value used for the generation of the test case.
\end{enumerate}

\begin{figure}[h!]
\centering
\subfloat[][d = 2]{
\includegraphics[width=0.5\textwidth]{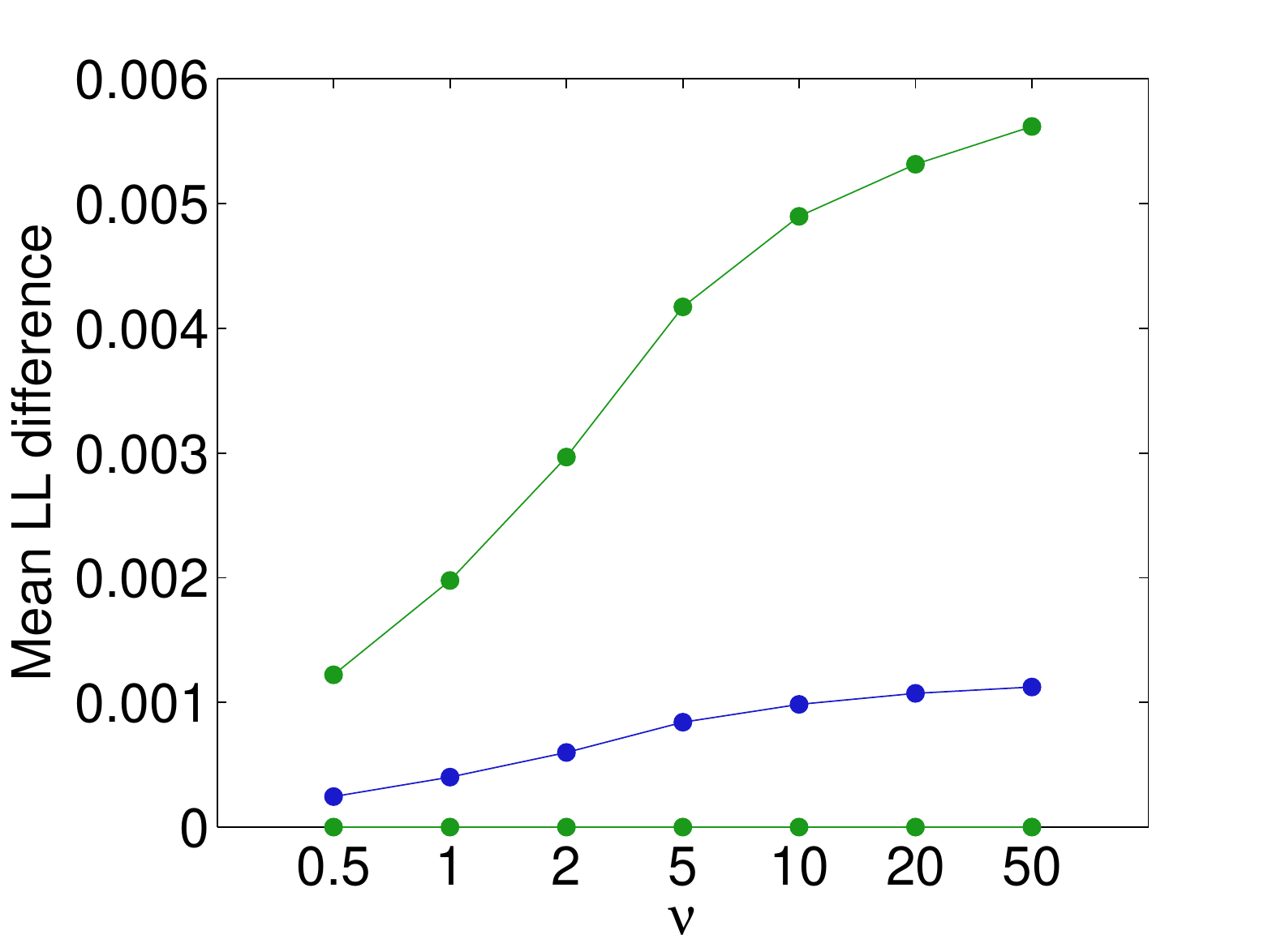}
\label{fig:subfig1}}
\subfloat[][d = 10]{
\includegraphics[width=0.5\textwidth]{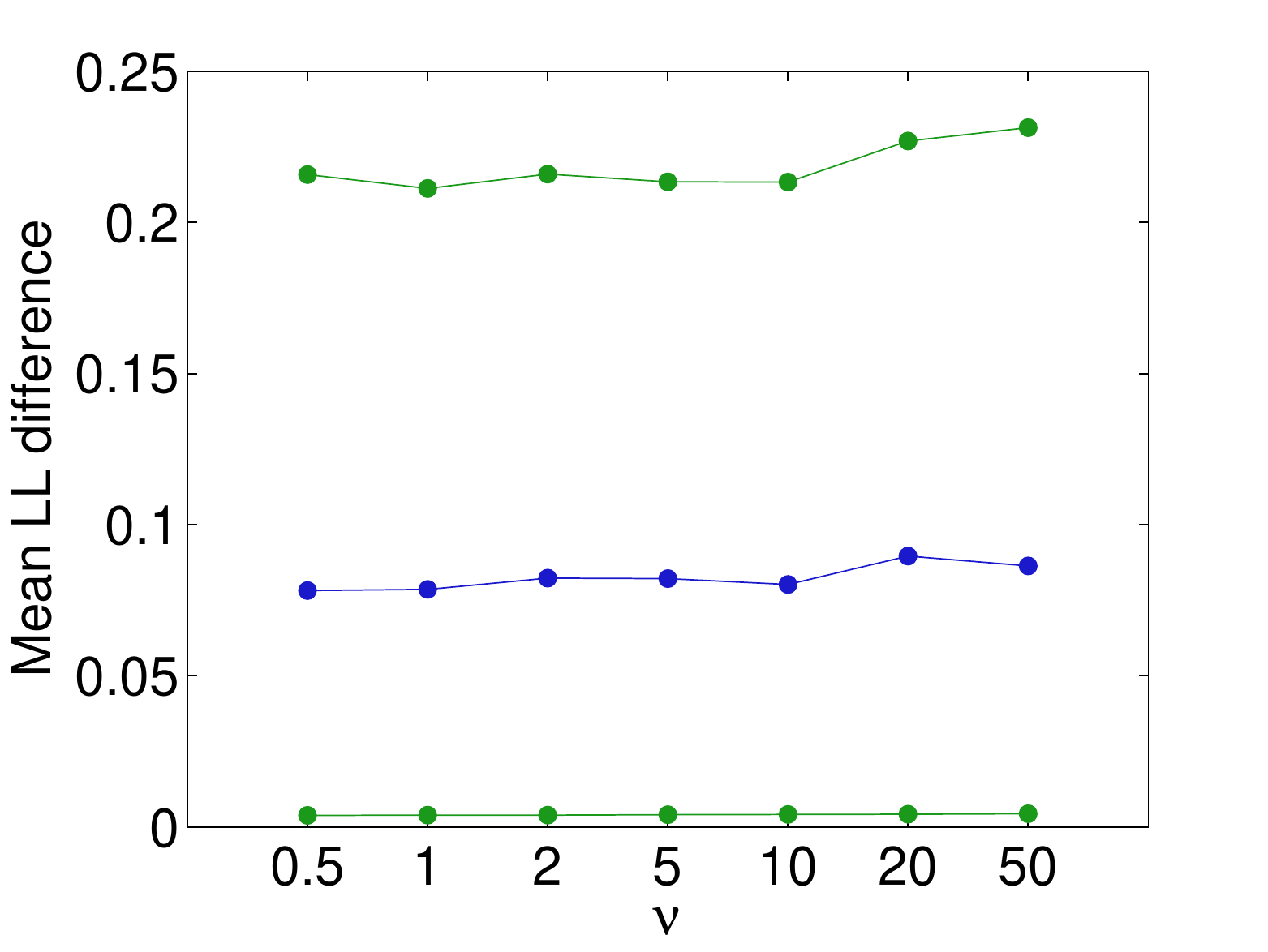}
\label{fig:subfig2}}
\qquad
\subfloat[][d = 25]{
\includegraphics[width=0.5\textwidth]{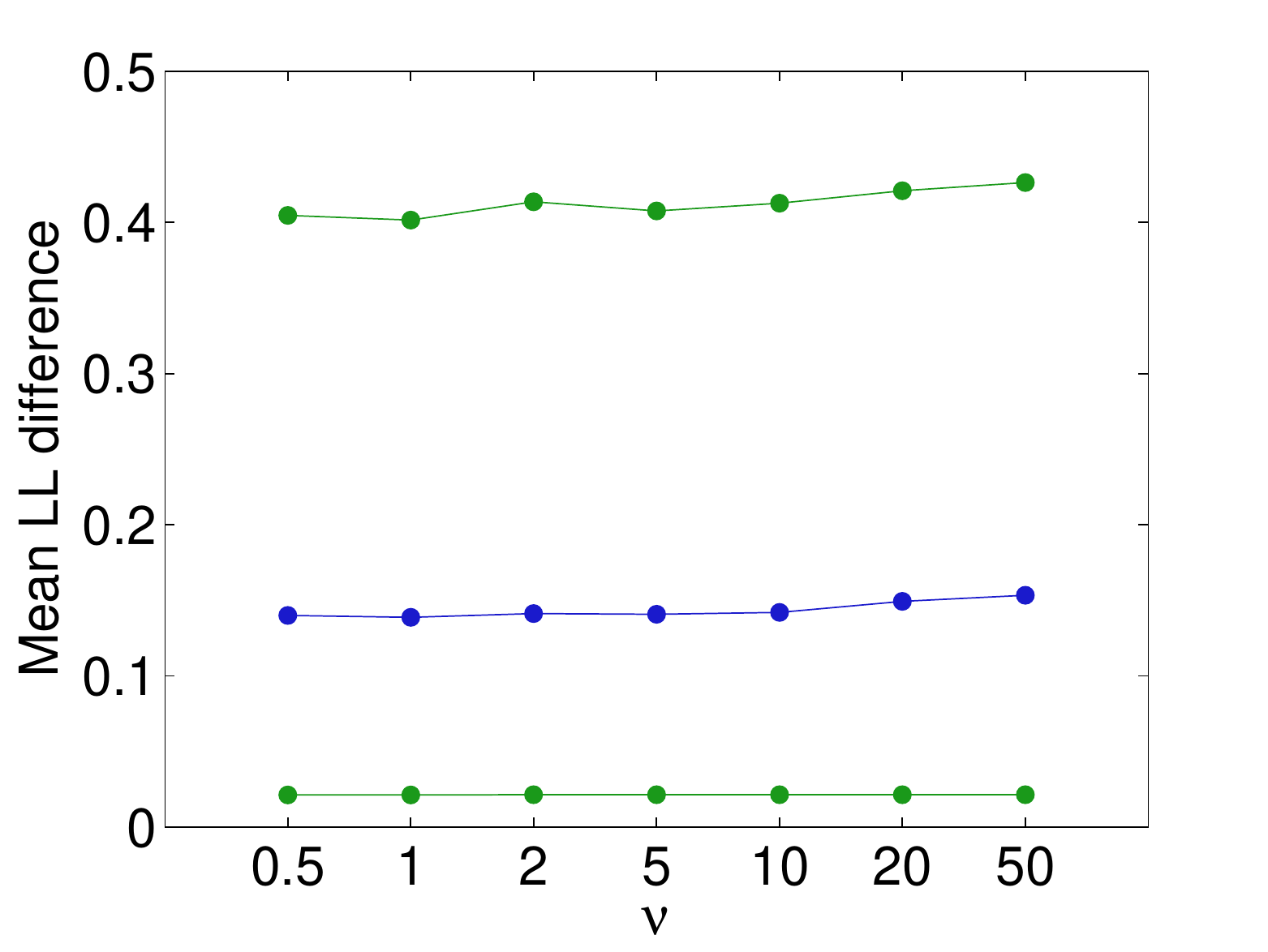}
\label{fig:subfig3}}
\caption{Mean (in blue) and 5-th and 95-th percentiles (in green) of the log-likelihood difference, normalized by the number of points (100), between the inverse gradient and the approximate method for different values of the dimension parameter, $d$.}
\label{fig:bouyeDeviation}
\end{figure}

\noindent In figure~\ref{fig:bouyeDeviation}, we plot the differences in the log-likelihood obtained with the inverse gradient and the approximate method for each value of the degrees-of-freedom parameter, which are all positive, as expected. In each case, we present the mean and 5-th and 95-th percentiles of the log-likelihood difference.\\

\noindent It is also worth mentioning that, although no convergence problems were observed in the case of 2 dimensions, the approximate method failed to converge in around $0.3\%$ of the cases for both 10 and 25 dimensions, while the inverse gradient converged in all of them.\\

\begin{figure}[h!]
\centering
\subfloat[][d = 2]{
\includegraphics[width=0.5\textwidth]{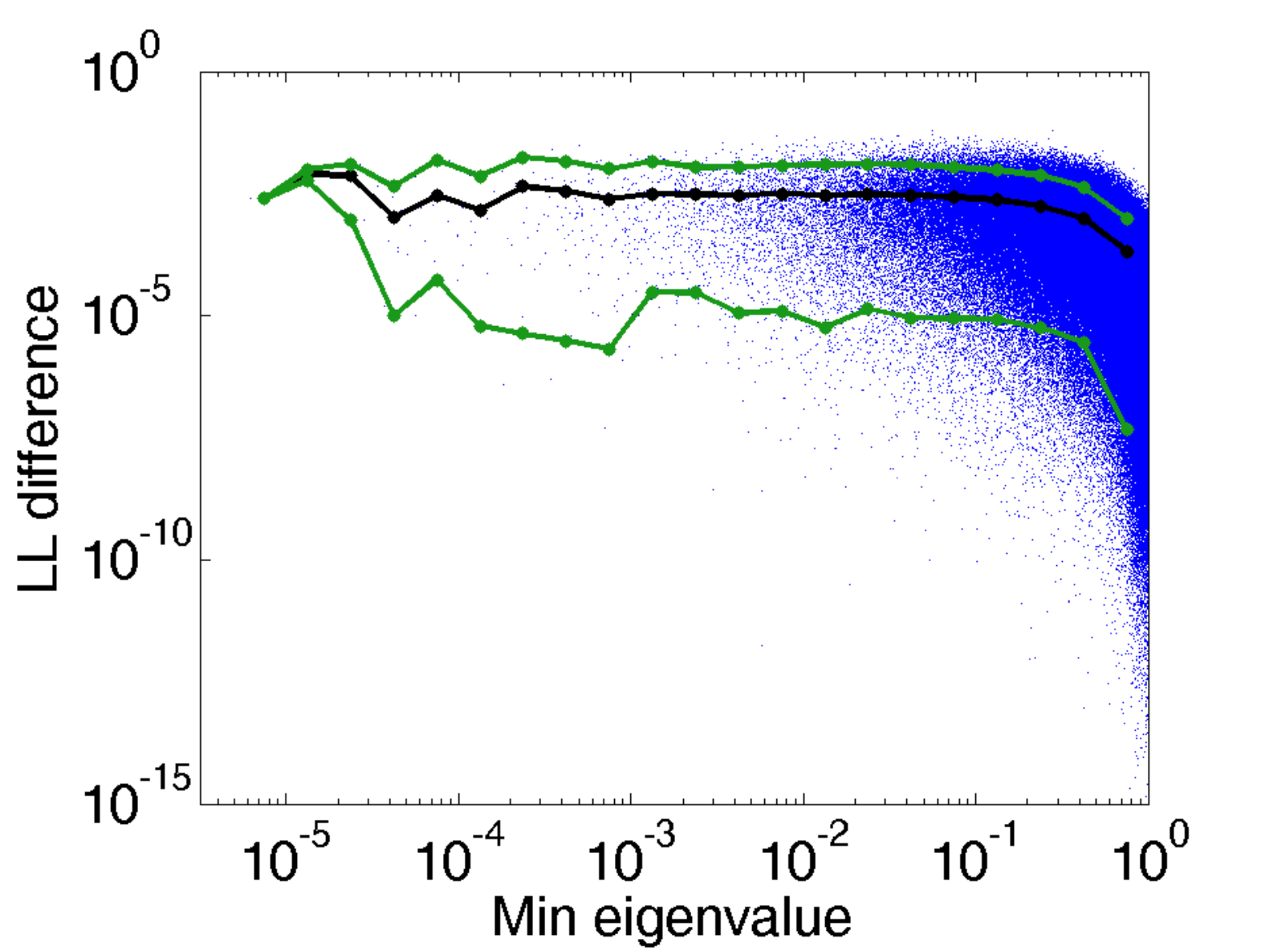}
\label{fig:subfig1}}
\subfloat[][d = 10]{
\includegraphics[width=0.5\textwidth]{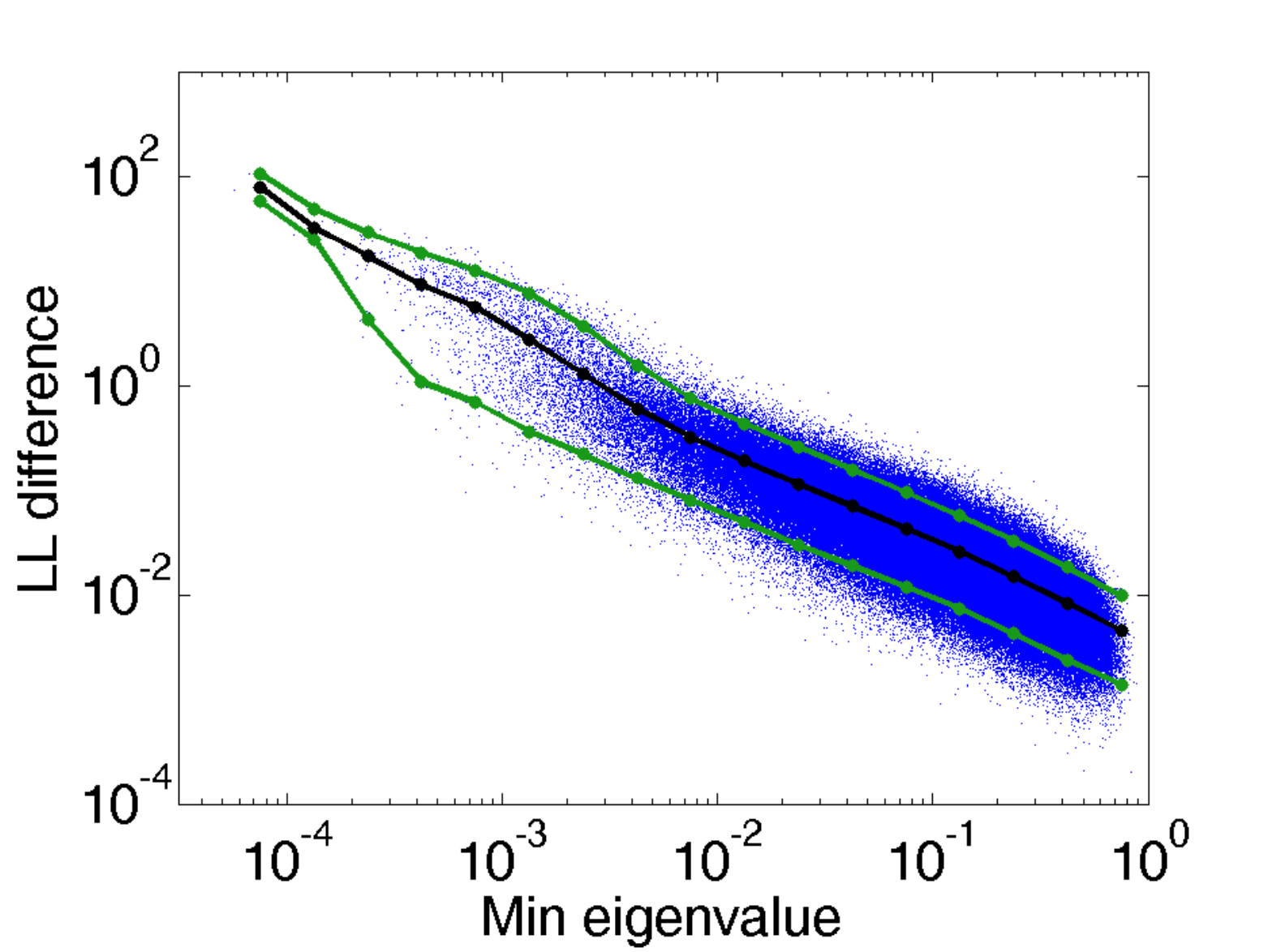}
\label{fig:subfig2}}
\qquad
\subfloat[][d = 25]{
\includegraphics[width=0.5\textwidth]{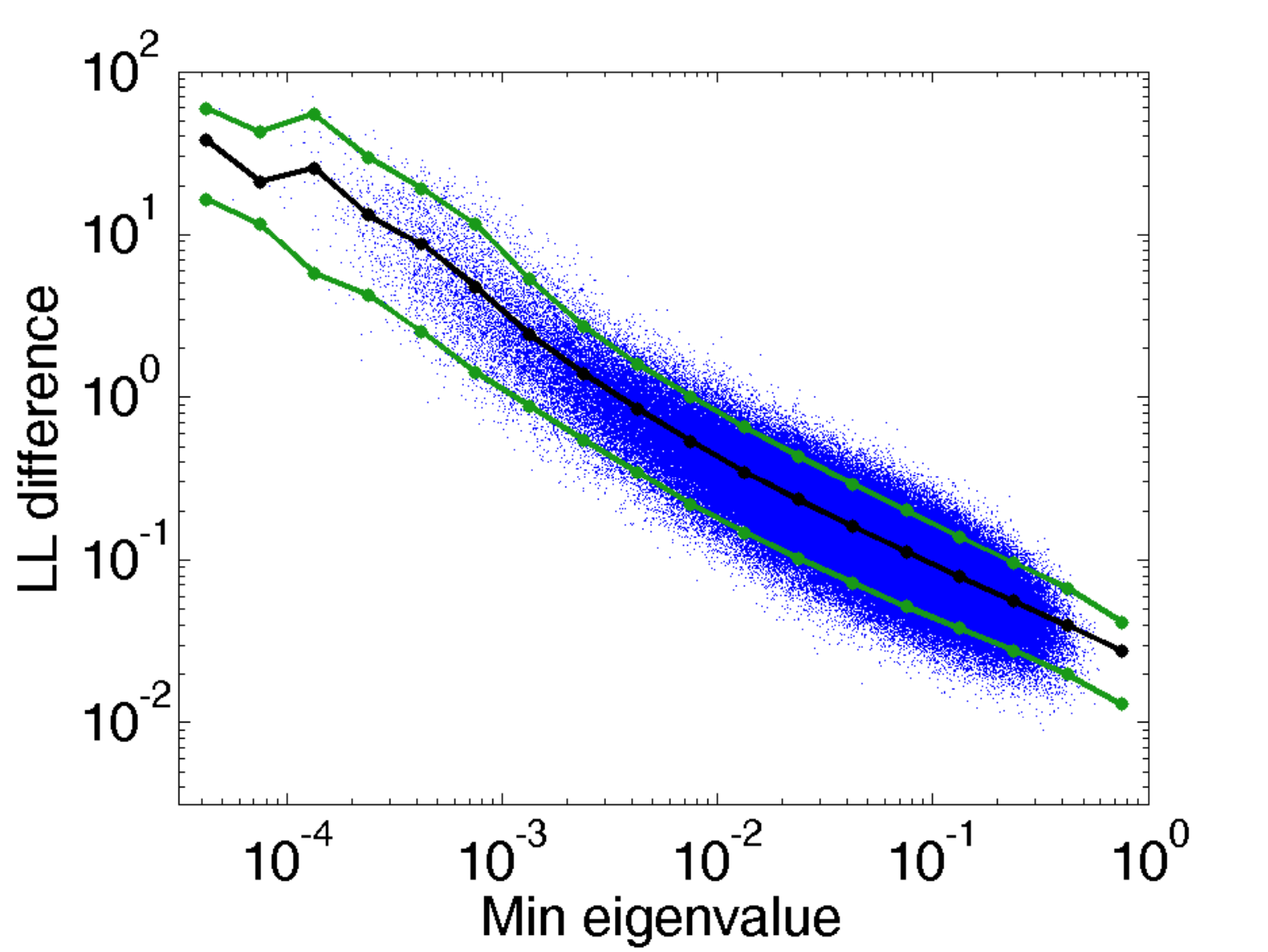}
\label{fig:subfig3}}
\caption{Scatter plot (in blue) of the log-likelihood deviation of the approximate method, normalized by the number of points (100), versus the minimum eigenvalue of the correlation matrix used to generate the samples, for different values of the dimension parameter, $d$. The lines represent the empirical mean (in black) and the 5-th and 95-th percentiles (in green) of the deviation for nearby values of the minimum eigenvalue.}
\label{fig:bouyeDevVsMinEigen}
\end{figure}

\noindent In figure~\ref{fig:bouyeDevVsMinEigen} we provide plots of the log-likelihood deviation of the approximate method as a function of the minimum eigenvalue of the correlation matrix used to generate the samples, for all the test cases generated of each dimension. Note that the deviations grow as the minimum eigenvalue becomes smaller, implying that the approximate method is especially unsuitable in cases where strong dependence is present.\\

\section{Conclusion}

\noindent We have presented an efficient procedure to obtain maximum likelihood estimates of the correlation parameter matrices of the Gaussian and Student's t copulas, which in principle can be extended to the family of elliptical copulas.\\

\noindent Other existing estimation procedures, extensively used in standard software packages, are either fast approximations or exact methods based on standard optimization techniques. The former do not return the true likelihood maximizers, and the computational time required by the latter grows with the number of parameters of the problem in such a way that these methods become inoperative for moderate numbers of dimensions.\\

\noindent The numerical tests performed for the Student's t distribution have shown that the log-likelihood gain obtained by using the proposed method instead of the approximation given in \cite{BDNRR00} increases with the dimension of the problem. This gain also increases when the minimum eigenvalue of the correlation matrix becomes smaller, showing a better behaviour of the presented algorithm for problems with strong dependence.\\

\noindent Numerical tests have been also used to compare the performance, in terms of computational time, of the proposed method versus exact methods provided by widely used software packages. The results have shown that the speed-up provided by the former is very significant, reaching several orders of magnitude for high-dimensional problems.\\

\textbf{Acknowledgements:} The authors thank Santiago Carrillo-Menéndez, Antonio Sánchez and 
Alberto Suárez for their valuable suggestions and corrections. 

\newpage 

\appendix

\section{Algorithm for the Gaussian and Student's t copulas}
\noindent In this section, the estimation algorithm is described in detail for the Gaussian and Student's t copulas. The starting point will be a data set of $n$ observations in the $d$-dimensional unit cube, $\{\mathbf{u}_t =(u_{t,1},...,u_{t,d}) \}_{t=1}^n, u_{t,i}\in(0,1)$. In the case of the Student's t copula, the value of the degrees-of-freedom parameter, $\nu$, is also given.
\begin{itemize}
    \item \textbf{Step 0}: Transform the observations using the inverse univariate distribution function
    \begin{eqnarray}
        g_{t,i} =& t_{\nu}^{-1}(u_{t,i}),  &\text{for the Student's t copula,}\nonumber\\
        s_{t,i} =& \Phi^{-1}(u_{t,i}),     &\text{for the Gaussian copula,}
    \end{eqnarray}
    and compute an initial estimate for the covariance matrix $\boldsymbol{\Sigma}_{[0]}$. In both cases, a good candidate for this initial seed is based on the approximation for the Gaussian copula 
    \begin{equation}
        \boldsymbol{\Sigma}_{[0]} = \frac{1}{n} \sum_{t=1}^n \textbf{g}_t  \textbf{g}_t^{\top}.
    \end{equation}        
    \item \textbf{Step 1}: Given the current estimate of the covariance matrix, $\boldsymbol{\Sigma}_{[m]}$, compute the projection to the correlation matrix space and find the inverse gradient direction 
    \begin{eqnarray}
        \mfr_{[m]} &=& \mf{A}_{[m]} \mfs_{[m]} \mf{A}_{[m]}, \nonumber\\
        \boldsymbol{\Delta}_{[m]} &=& -\frac{\partial L^*(\mfs_{[m]})}{\partial \mfs_{[m]}^{-1}}\nonumber\\
         &=& -\mf{A}_{[m]}^{-1} \Big( \mathcal{D}(\mfr_{[m]}) -  \mfr_{[m]} \ \text{diag} \left( \mathcal{D}(\mfr_{[m]}) \mfr_{[m]}^{-1} \right) \mfr_{[m]} \Big) \mf{A}_{[m]}^{-1} ,
    \end{eqnarray}
    with $(\mf{A}_{[m]})_{ij} = \frac{\delta_{ij}}{\sqrt{(\mfs_{[m]})_{ii}}}$. Below are the expressions for the log-likelihood derivative matrix $\mathcal{D}(\mfr) = \frac{\partial L(\mfr)}{\partial \mfr^{-1}}$ for the considered copulas.
    \begin{itemize}
        \item Gaussian copula:
        \begin{equation}
            \mathcal{D}(\mfr) = \frac{n}{2}\mfr - \frac{1}{2}\sum_{t=1}^n\mathbf{g}_t\mathbf{g}_t^{\top} .
        \end{equation}                
        \item Student's t copula:
        \begin{equation}
            \mathcal{D}(\mfr) = \frac{n}{2}\mfr - \frac{\nu+d}{2\nu}\sum_{t=1}^n\frac{\mathbf{s}_t\mathbf{s}_t^{\top}}{1+\frac{1}{\nu}\mathbf{s}_t^{\top} \mfr^{-1}\mathbf{s}_t}.
        \end{equation} 
    \end{itemize}
    \item \textbf{Step 2}: The next estimate, $\boldsymbol{\Sigma}_{[m+1]}$, is obtained by moving in the inverse gradient direction 
    \begin{equation}
        \mfs_{[m+1]} = \mfs_{[m]}+\lambda_{[m+1]}  \boldsymbol{\Delta}_{m} ,
    \end{equation}   
    where $\lambda_{[m]}$ is an adaptive step size. An initial step size $\lambda_{[0]}$ is chosen and, at each iteration, three step sizes are evaluated 
    \begin{equation}
        \lambda_{[m+1]} \in \{ k_1 \lambda_{[m]}, \lambda_{[m]}, k_2\lambda_{[m]}  \},
    \end{equation}
    where $0<k_1<1<k_2$, and the one with the highest log-likelihood is chosen if it fulfils two conditions:
    \begin{itemize}
        \item $\mfs_{[m+1]}$ is positive-definite.
        \item The log-likelihood increases: $L^*(\mfs_{[m+1]}) > L^*(\mfs_{[m]})$.
    \end{itemize} 
    If neither fulfils both conditions, the step size is reduced, $\lambda_{[m]}\longrightarrow k_1 \lambda_{[m]}$ and three new alternatives are evaluated.
    \item \textbf{Step 3}: If convergence in $\mfs_{[m]}$ has been achieved at step $m=M$, the projection to the correlation matrix space is returned
    \begin{equation}
        \widehat{\mfr} = \Pi(\mfs_{[M]} ) = \mf{A}_{[M]} \mfs_{[M]} \mf{A}_{[M]} 
    \end{equation}            
     and the algorithm terminates. Otherwise, repeat from step 1.
     
\end{itemize}

\noindent While better alternatives for this gradient algorithm can be probably constructed, we have found that the one presented yields a reasonably good performance with the choices 
\begin{equation}
    \lambda_{[0]}=\frac{1}{T}\ , \ k_1=\frac{1}{2}\ , \ k_2=\frac{4}{3} .
\end{equation}


\end{document}